\documentclass[twocolumn,prb,floatfix,citeautoscript,nofootinbib]{revtex4}                    
\usepackage{amsbsy}
\usepackage{latexsym,epsfig,graphicx}
\usepackage{dcolumn}
\usepackage{graphicx}
\usepackage{subfigure}
\usepackage{comment}
\usepackage{color}
\usepackage{bm}
\usepackage{mathrsfs}
\usepackage{amsfonts}
\usepackage{amsmath}
\usepackage{color}
\usepackage{amssymb}
\usepackage{xspace}
\usepackage{epstopdf}
\usepackage{tabularx}
\usepackage{longtable}
\usepackage[colorlinks=true, letterpaper=true, pdfstartview=FitV, linkcolor=blue, citecolor=blue, urlcolor=blue]{hyperref}
\usepackage[normalem]{ulem}

\begin{document}

\title{Magnon Nernst Effect in Magnon Spin Hall Systems}
\author{Yun-Mei Li}
\affiliation{Department of Physics, School of Physical Science and Technology, Xiamen University, Xiamen 361005, China}

\begin{abstract}
  Magnon spin Hall systems could hardly show experimentally observable particle and thermal transport phenomena intrinsically
  due to the spin cancellation. Here we demonstrated that the magnon spin Hall systems can exhibit magnon Nernst effect and thermal Hall effect under external magnetic field by considering two typical systems, i.e. the antiferromagnetically (AFM) coupled bilayer honeycomb ferromagnets and
  monolayer collinear honeycomb antiferromagnet. The both systems experience magnetic phase transitions from AFM phase to a field-polarized phase via a spin-flop (SF) phase or directly. In both systems, there exist magnon Nerst effect and also thermal Hall effect under a longitudinal temperature gradient, which can be regarded as the indicator of the magnetic phase transitions, with the Hall conductivity dependence on magnetic field consistent with the order of magnetic phase transition.
\end{abstract}

\maketitle

\section{Introduction}

In electronic systems, the transport properties can reveal the electronic structure information and also the topological properties.
One pronounced example is the Berry phase effect on the electrons~\cite{DXiao1}. The Berry curvature manifests
the electrons a transverse velocity, giving rise to various Hall effect, such as the anomalous Hall effect,
 topological Hall effect, spin Hall effect or valley Hall effect~\cite{DXiao1,NNagaosa,JSinova,DXiao2,DXiao3,JRSchaibley}.
 In a semiclassic interpretation, the Berry curvature
acts as a fictitious magnetic field in the momentum space, and the related effective Lorentz force bends the electron trajectory to the transverse direction.
The observation of these effects usually requires time-reversal or inversion symmetry breaking.

The concepts of Berry phase and Berry curvature are not restricted on the electronic systems, but also applies
to the bosons, such as phonon~\cite{LZhang,TQin,TSaito}, photons~\cite{RYChiao,FDMHaldane,ZWang}
and also magnetic excitations, or magnons~\cite{HKatsura,YOnose,SFujimoto,RMatsumoto,YSLu,RRNeumann,AMook1,AMook3,
RMatsumoto2,PMGunnink,PMellado,AOkamoto,KAHoogdalem,SAOwerre,RCheng,VAZyuzin,YMLi2},
inducing
the bosonic analog of anomalous Hall effects.
In this paper, we focus on the magnons as they are easily manipulated by external magnetic field.
Meanwhile, the external magnetic field may drive the system into different magnetic phases, giving distinct
Berry curvature distribution in bands or topological phases, thus showing different transport phenomena.
For magnons, the Berry curvature mainly comes from the Dzyaloshinskii-Moriya interaction
(DMI)~\cite{HKatsura,YOnose,AMook1,AMook3,YSLu}, dipolar interactions~\cite{RMatsumoto2,PMGunnink,PMellado,AOkamoto},
the magnetic textures~\cite{KAHoogdalem,SFujimoto,SAOwerre}, or the antiferromagnetic (AFM) coupling~\cite{RCheng,VAZyuzin,RRNeumann}.
As magnon is charge neutral, it is hard to observe the Hall effect directly. But magnons can carry heat current
and thus exhibit a thermal Hall effect, which is experimentally observable.
Beisdes the magnon Hall effect, the antiferromagnetic order endows effective magnon spins, giving rise to magnonic analog of spin Hall effect
when the two spin branches host opposite Berry curvature~\cite{RCheng,VAZyuzin,YMLi,HKondo1}.
Same to the electronic systems, the magnon spin Hall systems exhibit zero net magnon current in the transverse direction,
thus hold a vanishing thermal Hall effect, making it hard to detect experimentally.

In this paper, we demonstrated that the magnon spin Hall systems can exhibit a magnon Nernst (Hall) and thermal Hall effects under external magnetic field.
We here consider two typical systems showing magnon spin Hall effect in the absence of magnetic field.
The first one is the bilayer honeycomb ferromagnets with AFM interlayer interactions.
The monolayer one with next-nearest-neighboring DMI is quite widely investigated, which show magnon anomalous Hall effect.
The second system is monolayer collinear honeycomb antiferromagnet
with NNN DMI. Both of the systems host magnon spin Hall effect due to the AFM coupling and NNN DMI.
Under magnetic field, both of the systems experiences magnetic phase transitions from
the AFM ordering between layers or sublattices to a field-polarized (FP) phase via a first-order spin-flop (SF) transition or directly at critical values
depending on the easy-axis magnetic anisotropy. In all the phases, the topology of the magnon bands
are characterized by the Chern number or spin Chern number.
In the AFM phase, the broken degeneracy of magnon bands give a magnon Hall effect, manifest it the Nernst effect under the temperature gradient,
with the Nernst conductivity depending on the magnetic field.
The conductivity also experiences a first-order transition when the both systems go into
the SF phase or the FP phase from AFM phase.
From the SF phase to the FP phase,
the conductivity change continuously but not smoothly, consistent with the phase transition.
The thermal Hall effect of magnons show the same behavior to the Nernst effect.
The different behavior of Nernst effect and thermal Hall effect of magnons
indicates that we can use them as the indicator of the magnetic phase transitions in magnon spin Hall systems.
We further find the other magnon spin Hall systems have direct correspondence to the two typical systems in phase
diagram and transport properties.

This paper is organized as follows. In Sec. II, we present the theoretical models, the method to determine the magnetic phase transition,
calculation of the magnon bands and Nernst and thermal Hall conductivity. In Sec. III, we give the phase diagram, the magnon bands
in different magnetic phases and the band topology. The Nernst effect and thermal Hall effect dependence on the magnetic field are also discussed.
Finally, we summarize in Sec. IV.

\section{Theoretical models}

\begin{figure}[t]
  \centering
  \includegraphics[width=0.5\textwidth]{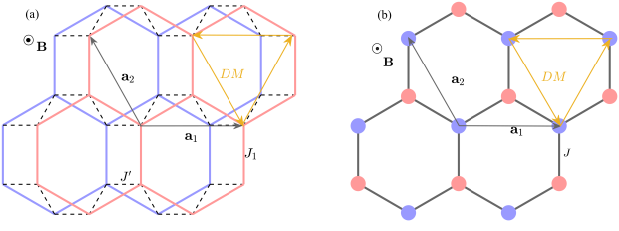}\\
  \caption{Two typical magnon spin Hall systems. (a) monoclinic stacked bilayer honeycomb ferromagnets.
  The dashed black lines denote the interlayer antiferromagnetic coupling.
  The blue lines denote the ground state pointing up layer while the red pointing down. The green lines denote the DMI.
  (b) the monolayer honeycomb antiferromagnet with the color notation same to (a).}\label{fig1}
\end{figure}

Here we consider two typical magnon spin Hall systems, as shown in Fig.~\ref{fig1}.
The first one is the bilayer honeycomb ferromagnets with AFM interlayer coupling, which have been proved
to host a $Z_{2}$ topological invariant~\cite{HKondo1,YMLi}. Without the Pauli exclusion, the boson usually can not give a quantized
response to the external field, the system would thus give a spin Hall effect.
Here we take the monoclinic stacked bilayer CrI$_{3}$ as the example material~\cite{BHuang,NSivadas,PJiang}, illustrated in Fig.~\ref{fig1} (a).
We apply a perpendicular magnetic field $\mathbf{B}=B\mathbf{e}_{z}$.
The spin interaction Hamiltonian for the bilayer honeycomb ferromagnets is given by
\begin{align}
  H_{1}=&-\sum_{ i\neq j,l}J_{ij}\mathbf{S}_{i,l}\cdot\mathbf{S}_{j,l}+D\sum_{\langle\langle ij\rangle\rangle,l}\nu_{ij}\hat{z}\cdot(\mathbf{S}_{i,l}\times\mathbf{S}_{j,l})       \nonumber \\
   &-\sum_{i,l}[hS_{i,l}^{z}+K_{z}(S_{i,l}^{z})^2] +\sum_{\langle ij\rangle}J^{\prime}\mathbf{S}_{i,1}\cdot\mathbf{S}_{j,2},   \label{eq1}
\end{align}
where $h=g\mu_{B}B$, $\mu_{B}$ the Bohr magneton, $g$ denotes the g-factor. $l=1,2$ is the layer index. $J_{ij}>0$ characterizes intralayer FM coupling.
$D$ represents the next-nearest-neighboring (NNN) Dzyaloshinskii-Moriya interaction (DMI) strength.
$\nu_{ij}=\pm 1$ with $+$ ($-$) for counterclockwise (clockwise) circulation.
$K_{z}$ characterizes the single-ion easy-axis anisotropy. The last term denote the AFM interlayer coupling with $J^{\prime}>0$.

The second system is the monolayer honeycomb antiferromagnet,
demonstrated before to exhibit spin Nernst (Hall) effect~\cite{RCheng,VAZyuzin}, illustrated in Fig.~\ref{fig1} (b).
For now, we did not find the experimentally demonstrated candidate materials.
The spin interaction Hamiltonian
\begin{align}
  H_{2}=&J\sum_{\langle ij\rangle}\mathbf{S}_{i}\cdot\mathbf{S}_{j}+D\sum_{\langle\langle ij\rangle\rangle}\nu_{ij}\hat{z}\cdot(\mathbf{S}_{i}\times\mathbf{S}_{j})  \nonumber \\
  &-\sum_{i}[hS_{i}^{z}+K_{z}(S_{i}^{z})^2].   \label{eq2}
\end{align}
Here $J>0$ characterize the NN antiferromagnetic coupling. The same parameter denotes the same meaning to the first systems.

Under the external  magnetic field, the AFM ordering between layers or sublattices will not always be the ground state and the
both systems will experience magnetic phase transitions.
We can restrict the transition in the $xz$ plane due to the anisotropic exchange interactions.
The ground state magnetic ordering can be
obtained by minimizing the interaction energies.
The spins in the same layer for the first system and in the same sublattice for the second system will align to the same direction,
as the NNN DMI does not compromise the ferromagnetic ground state~\cite{AMook2}.
For the two systems, the interaction energy per unit cell can be written in a unified form,
$\overline{E}= NJ_{AFM}S^{2}\delta\cos(\theta_{1}-\theta_{2})-NhS(\sin\theta_{1}+\sin\theta_{2})-NK_{z}S^{2}(\sin^{2}\theta_{1}+\sin^{2}\theta_{2})$,
where $N=2$ for the first system and $N=1$ for the second system. $J_{AFM}$ is the AFM coupling strength, $S$ the spin value. $\delta=2$ for the first system and $\delta=3$ for the second systems. $\theta_{1}$ and $\theta_{2}$ the rotation angle of spins on the separate layer or sublattice with respect to positive $x$-axis.

The magnons are the spin excitations from the magnetic ground state. To get the dispersion relation,
we apply the linear spin wave theory.
We can write the spin vector as $\mathbf{S}_{i}=S_{i}^{r}\mathbf{e}_{i}^{r}+S_{i}^{\theta}\mathbf{e}_{i}^{\theta}+S_{i}^{\varphi}\mathbf{e}_{i}^{\varphi}$
on the sphere and apply the Holstein-Primakoff (HP) transformation,
$S_{i}^{r}=S-a_{i}^{\dagger}a_{i}$, $S_{i}^{+}=\sqrt{2S}a_{i}$, $S_{i}^{-}=\sqrt{2S}a_{i}^{\dagger}$,
$S_{i}^{\pm}=S_{i}^{\theta}\pm iS_{i}^{\varphi}$. Then we make the Fourier transformation
$a_{i}=\frac{1}{N}\sum_{\mathbf{k}}e^{i\mathbf{k}\cdot\mathbf{r}_{i}}b_{\alpha,\mathbf{k}}$,
$a_{i}^{\dagger}=\frac{1}{N}\sum_{\mathbf{k}}e^{-i\mathbf{k}\cdot\mathbf{r}_{i}}b_{\alpha,\mathbf{k}}^{\dagger}$ ($\alpha$ is the sublattice index in a unit cell).
The single-particle term vanished by applying the ground state solution. We here neglect the three and more particle term, and
the Hamiltonian is expressed as $H=\frac{1}{2}\sum_{\mathbf{k}}\Psi_{\mathbf{k}}^{\dagger}H_{\mathbf{k}}\Psi_{\mathbf{k}}$ in the basis
$\Psi_{\mathbf{k}}=(\psi_{\mathbf{k}},\psi_{-\mathbf{k}}^{\dagger})^{T}$ up to the quadratic term.
$\psi_{k}=(b_{A,1,\mathbf{k}},b_{B,1,\mathbf{k}},b_{A,2,\mathbf{k}},b_{B,2,\mathbf{k}})^{T}$ for the first system and
$\psi_{k}=(b_{A,\mathbf{k}},b_{B,\mathbf{k}})^{T}$ for the second system. The form of the Hamiltonian
$H_{\mathbf{k}}$ depends on the system and the magnetic phases. We will list them in the next section.
Diagonalizing the quadratic form invokes the Bogoliubov transformation,
$T_{\mathbf{k}}^{\dagger}H_{\mathbf{k}}T_{\mathbf{k}}=diag\{E_{\mathbf{k}},E_{-\mathbf{k}}\}$, with paraunitary eigenvectors satisfying
$T_{\mathbf{k}}^{\dagger}\tau_{z}T_{\mathbf{k}}=\tau_{z}$ and $\tau_{z}$ is the Pauli matrix acting on the particle-hole space~\cite{RShindou}.
For magnon, we usually adopt $E_{\mathbf{k}}=\hbar\omega_{\mathbf{k}}$ with $\omega_{\mathbf{k}}$ the frequency of spin wave modes.

To describe the transport properties of magnons, the Berry curvature is essential.
We introduce the Berry connection ${A}_{\mu}^{n}(\mathbf{k})=i\mathrm{Tr}[\Gamma^{n}\tau_{z}T_{\mathbf{k}}^{\dagger}\tau_{z}(\partial_{k_{\mu}}T_{\mathbf{k}})]$
and the Berry curvature $\bm\Omega_{n}=\nabla_{\mathbf{k}}\times\mathbf{A}^{n}(\mathbf{k})$ for $n$-th magnon band,
where $\Gamma^{n}$ is the diagonal matrix taking $+1$ for n-th digaonal component and zero otherwise.
Due to the external magnetic field, the topology of magnon bands are described by the Chern number, given by
$C_{n}=\frac{1}{2\pi}\int_{BZ} \Omega_{n}^{z}(\mathbf{k}) d^{2}\mathbf{k}$ for $n$-th band.

Here we are mainly interested in the transport properties induced by the magnons.
The non-vanishing Berry curvature of the magnon bands indicate that magnon wavepacket will experience
an additional force during motion, giving rise to Hall-like effect and also thermal Hall effect under a thermal gradient.
The magnon Hall current is given by~\cite{RMatsumoto}
\begin{equation*}
  \mathbf{j}=-\frac{k_{B}}{\hbar}\hat{z}\times\nabla T\sum_{n}\int[d\mathbf{k}]\Omega_{n}^{z}(\mathbf{k})c_{1}(\rho_{n,\mathbf{k}}) = \alpha_{xy}\hat{z}\times\nabla T,
\end{equation*}
where $[d\mathbf{k}]=d^{2}\mathbf{k}/(2\pi)^{2}$, $k_{B}$ is the Boltzmann constant, $\hbar$ reduced Planck constant, $c_{1}(x)=(1+x)\log(1+x)-x\log x$ and $\rho_{n,\mathbf{k}}=(e^{\beta E_{n,\mathbf{k}}}-1)^{-1}$ is the Bose distribution function, $\beta=\frac{1}{k_{B}T}$, $T$ is the temperature.
The Nernst conductivity
\begin{equation}
  \alpha_{xy}=-\frac{k_{B}}{\hbar}\sum_{n}\int[d\mathbf{k}]\Omega_{n}^{z}(\mathbf{k})c_{1}(\rho_{n,\mathbf{k}}),  \label{eq3}
\end{equation}
and the Nernst coefficient $\mathcal{N}=\alpha_{xy}/B$. The motion of magnon would also give a thermal Hall effect, with the thermal Hall current given by
\begin{equation*}
  \mathbf{j}_{Q}= -\frac{k_{B}^{2}T}{\hbar}\hat{z}\times\nabla T\sum_{n}\int[d\mathbf{k}]\Omega_{n}^{z}(\mathbf{k})c_{2}(\rho_{n,\mathbf{k}})=\kappa_{xy}\hat{z}\times\nabla T,
\end{equation*}
and the thermal Hall conductivity~\cite{RMatsumoto,HKatsura}
\begin{equation}
  \kappa_{xy}=-\frac{k_{B}^{2}T}{\hbar}\sum_{n}\int[d\mathbf{k}]\Omega_{n}^{z}(\mathbf{k})c_{2}(\rho_{n,\mathbf{k}}).  \label{eq4}
\end{equation}
with $c_{2}(x)=(1+x)(\log\frac{1+x}{x})^{2}-(\log x)^{2}-2Li_{2}(-x)$, where $Li_{2}(x)$ is the polylogarithm function.

\section{Results and discussions}

\subsection{Phase diagram and magnon bands}

\begin{figure}[t]
  \flushleft
  \includegraphics[width=0.5\textwidth]{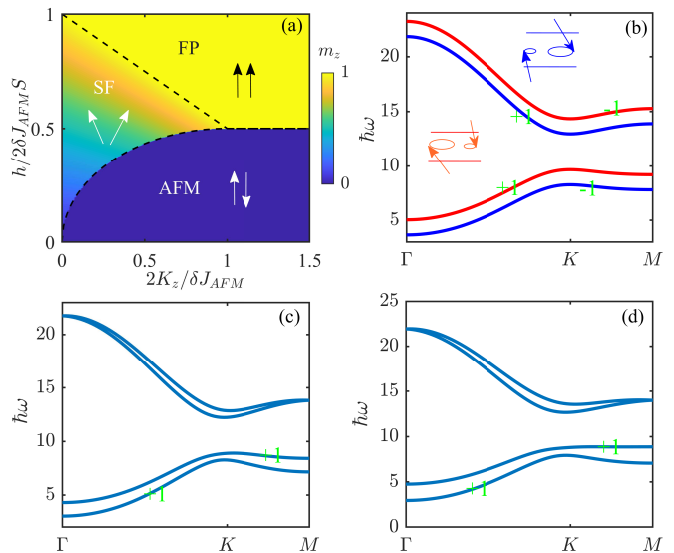}\\
  \caption{The phase diagram and magnon bands at three magnetic phases for the first magnon spin Hall system,
  i.e. AFM coupled bilayer honeycomb ferromagnets. (a) the phase diagram.
  The dashed black lines denote the phase transition critical value. The color denotes the order parameter, i.e. magnetization.
  (a-d). The magnon bands for first system at $B=4$ T (b), $B=5.5$ T (c) and $B=7$ T. Here we take the bilayer CrI$_{3}$
  as the candidate material. $S=\frac{3}{2}$, $J_{1}=2.01$ meV, $J_{2}=0.16$ meV, $J^{\prime}=0.3$ meV,
  $D=0.3$ meV, $K_{z}=0.22$ meV, $g\simeq 3$~\cite{LChen1,LChen2}.We labelled the Chern number of the corresponding bands
  by green numbers.}\label{fig2}
\end{figure}

The ground state magnetic ordering between layer and sublattice depends on the external magnetic field and the magnetic interaction parameters.
We plot the phase diagrams in Fig.~\ref{fig2} (a) and also the order parameter, i.e. the magnetization.
The two systems share a similar phase transition diagram.
There are two distinct magnetic phase transition
paths depending on the easy-axis anisotropy $K_{z}$.
Firstly, when $K_{z}<\frac{\delta}{2}J_{AFM}$, the magnetic order stays in the
AFM phase below a critical value $B_{c1}=2S\sqrt{K_{z}(\delta J_{AFM}-K_{z})}/g\mu_{B}$.
Further increasing the magnetic field, the both systems experience a spin-flop (SF) transition.
In this phase, the spins are partly aligned to the magnetic field direction and $\theta_{2}=\pi-\theta_{1}$. The magnetic phase transition
is a first-order between AFM and SF phases, with a jump in the order parameter $m_{z}$, as shown in Fig.~\ref{fig2} (a).
The angle $\theta_{1}=\sin^{-1}\frac{h}{2S(\delta J_{AFM}-K_{z})}$. Increasing the magnetic field
to another critical value $B_{c2}=2S(\delta J_{AFM}-K_{z})/g\mu_{B}$, $\theta_{1}=\theta_{2}=\pi/2$,
the spins are fully polarized along the magnetic field. A larger magnetic field will not affect the ground state magnetic order.
This transition is a second-order phase transition.
Secondly, when $K_{z}>\frac{\delta}{2}J_{AFM}$, the both systems will go into the FP phase directly from the AFM phase
with the critical value $B_{c3}=\delta J_{AFM}S/g\mu_{B}$ and the phase transition is in first-order.

In the three magnetic phases, the systems hold different ground state spin configurations.
The spin excitations are expected to be different.
We here discuss the magnon Hamiltonians and the topological phases in different magnetic phases.
We first discuss the bilayer honeycomb ferromagnets. The magnon Hamiltonian in the AFM phase is given by
\begin{equation}\label{eq5}
  H_{1,\mathbf{k}}^{A}=h_{1,0}^{A}+\left(\begin{array}{cccc}
  h_{1,\mathbf{k}}^{A}& 0 & 0 & h_{\perp,\mathbf{k}} \\
  0 & h_{1,-\mathbf{k}}^{A*} & h_{\perp,\mathbf{k}}^{\dagger} & 0 \\
  0 & h_{\perp,\mathbf{k}} & h_{1,-\mathbf{k}}^{A*} & 0 \\
  h_{\perp,\mathbf{k}}^{\dagger} & 0 & 0 &  h_{1,\mathbf{k}}^{A}
  \end{array}\right),
\end{equation}
where $h_{1,0}^{A}=(2K_{z}+3J_{1}+2J^{\prime}+2J_{2}p_{\mathbf{k}})S+h\tau_{0}s_{z}\sigma_{0}$,
$h_{1,\mathbf{k}}^{A}=\mathbf{h}\cdot\bm\sigma$ with
$h_{x}=-J_{1}S\mathrm{Re}(f_{\mathbf{k}})$, $h_{y}=J_{1}S\mathrm{Im}(f_{\mathbf{k}})$,
$f_{\mathbf{k}}=1+e^{-i\mathbf{k}\cdot\mathbf{a}_{3}}+e^{i\mathbf{k}\cdot\mathbf{a}_{2}}$,
$h_{z}=-2DSg(\mathbf{k})$, $g(\mathbf{k})=\sum_{i=1}^{3}\sin(\mathbf{k}\cdot\mathbf{a}_{i})$,
$p_{\mathbf{k}}=\sum_{i=1}^{3}\cos(\mathbf{k}\cdot\mathbf{a}_{i})$,
$\mathbf{a}_{1}=(\sqrt{3},0,0)$, $\mathbf{a}_{2}=(-\frac{\sqrt{3}}{2},\frac{3}{2},0)$, $\mathbf{a}_{3}=-(\mathbf{a}_{1}+\mathbf{a}_{2})$ intralyer NNN vectors. $\bm\tau$, $\mathbf{s}$, $\bm\sigma$ are pauli matrices acting on particle-hole, layer and sublattice index, respectively.
\begin{equation*}
  h_{\perp,\mathbf{k}}=J^{\prime}S\left(\begin{array}{cc}
  1 &    e^{i\mathbf{k}\cdot\mathbf{a}_{3}} \\
   e^{i\mathbf{k}\cdot\mathbf{a}_{2}}  & 1
  \end{array}\right).
\end{equation*}
In the absence of magnetic field, $h=0$, the Hamiltonian in Eq.(\ref{eq5}) hold a nontrivial $Z_{2}$ topological phase and the magnon bands are doubly degenerate.
The magnetic field breaks the $Z_{2}$ topology and lift the band degeneracy, as shown in Fig.~\ref{fig2} (b).
We can decompose the Hamiltonian into two independent sectors
$H_{1,\mathbf{k}}^{A}=H_{1,\uparrow}^{A}\oplus H_{1,\downarrow}^{A}$, labelled by the (pseudo-)spin index, arising from the
AFM interlayer coupling.
The split bands are spin polarized and we plot the illustration of magnon modes of opposite spins in the insets in Fig.~\ref{fig2} (b).
In the $Z_{2}$ broken phase, we can use the spin Chern number~\cite{LSheng,DNSheng,EProdan,YYang} as the topological invariant $C_{\pm}=\pm 1$ as we labelled in Fig.~\ref{fig2} (b).

In the SF phase, the magnon Hamiltonian
\begin{equation}\label{eq6}
  H_{1,\mathbf{k}}^{S}=h_{1,0}^{S}+\left(\begin{array}{cccc}
  h_{1,\mathbf{k}}^{S}& h_{12,\mathbf{k}}^{1,S} & h_{11}^{S} & h_{\perp,\mathbf{k}}^{S} \\
   & h_{1,\mathbf{k}}^{S} & h_{\perp,\mathbf{k}}^{S\dagger} & h_{11}^{S} \\
   &  & h_{1,-\mathbf{k}}^{S*} & h_{12,\mathbf{k}}^{1,S} \\
  H.c. &  &  &  h_{1,-\mathbf{k}}^{S*}
  \end{array}\right).
\end{equation}
$h_{1,0}^{S}=[K_{z}(2\sin^{2}\theta-\frac{1}{2}\cos^{2}\theta)+3J_{1}+2J^{\prime}\cos2\theta+2J_{2}p_{\mathbf{k}}]S+h\sin\theta$,
$h_{12,\mathbf{k}}^{1,S}=\sin^{2}\theta h_{\perp,\mathbf{k}}$, $h_{11}^{S}=-\cos^{2}\theta\frac{K_{z}S}{2}$,
$h_{\perp,\mathbf{k}}^{S}=-\cos^{2}\theta h_{\perp,\mathbf{k}}$,
$h_{1,\mathbf{k}}^{S}=h_{x}\sigma_{x}+h_{y}\sigma_{y}+\sin\theta h_{z}\sigma_{z}$,
$\sin\theta=\frac{h}{2S(2J^{\prime}-K_{z})}$. The topology of the magnon bands are determined by $h_{1,\mathbf{k}}^{S}$
with the Chern  number labelled in Fig.~\ref{fig2} (c). The magnon band of FP phase is the same to Eq.~(\ref{eq6})
by setting $\theta=\pi/2$ and keeping only the particle branch with the bands shown in Fig.~\ref{fig2} (d).
In the SF and FP phase, we get topological magnon bandgap with high Chern number ($C>1$). This will largely enlarge
the Nernst conductivity and thermal Hall conductivity, discussed subsequently.

\begin{figure}[t]
  \centering
  \includegraphics[width=0.5\textwidth]{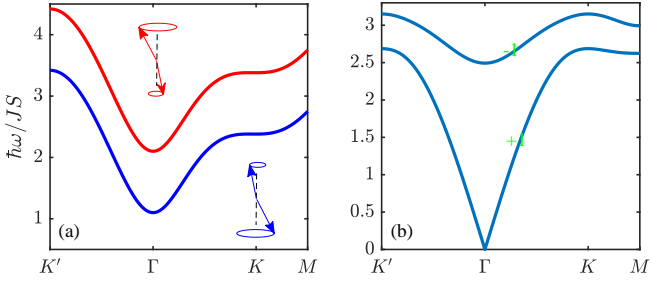}\\
  \caption{The magnon bands for the second systems at the AFM (in (a)) and SF (in (b)) phase.
  The parameters are adopted as follows: $D=0.1J$, $K_{z}=0.2J$, $h=0.5JS$ in (a) and $h=2.5JS$ in (b).}\label{fig3}
\end{figure}

For the monolayer honeycomb antiferromagnet, the magnon Hamiltonian in the AFM phase
is given by
\begin{equation}\label{eq7}
  H_{2,\mathbf{k}}^{A}=h_{2,0}^{A}+\left(
  \begin{array}{cccc}
  h_{z,\mathbf{k}} & 0 & 0 & JSf_{\mathbf{k}} \\
  0 & h_{z,\mathbf{k}} & JSf_{-\mathbf{k}} & 0\\
  0 & JSf_{\mathbf{k}} & h_{z,-\mathbf{k}} & 0\\
  JSf_{-\mathbf{k}} & 0 &0 & h_{z,-\mathbf{k}}
  \end{array}
  \right),
\end{equation}
where $h_{2,0}^{A}=3JS+2K_{z}S+h\tau_{0}\sigma_{z}$. This Hamiltonian can be recast into two sectors even at finite $h$.
As demonstrated in former work~\cite{RCheng,VAZyuzin}, the two sectors give us two magnon branches, denoted as
the magnon spin, $H_{2,\mathbf{k}}^{A}=H_{2,\uparrow}^{A}\oplus H_{2,\downarrow}^{A}$.
Without magnetic field, the two magnon branches are degenerate. The magnetic field breaks the degeneracy, as shown in Fig~\ref{fig3} (a).
Here we should notice that for each band, the Berry curvature hold the relation, $\bm\Omega_{\mathbf{k}}=-\bm\Omega_{-\mathbf{k}}$,
the Chern number of both bands are zero. The NNN DMI brings the asymmetry magnon bands, leading to nonreciprocal transport behavior
of magnons. As a result, for each spin-polarized band, the Berry curvature effect does not vanish, bringing the spin Nernst effect, demonstrated in
previous works~\cite{RCheng,VAZyuzin}. Here the split bands by magnetic field will induce imbalanced spin population at finite temperatures, expected to
give rise to nonvanishing Nernst effect.

In the SF phase, the magnon Hamiltonian is in the following form,
\begin{equation}\label{eq8}
  H_{2,\mathbf{k}}^{S}=h_{2,0}^{S}+\left(\begin{array}{cccc}
  h_{z,\mathbf{k}}^{\prime} & h_{12,\mathbf{k}}^{2,S} & \Delta_{11}^{S}  & \Delta_{12,\mathbf{k}}^{S}  \\
     & -h_{z,\mathbf{k}}^{\prime} & \Delta_{12,-\mathbf{k}}^{S} & \Delta_{11}^{S} \\
    &  &   h_{z,-\mathbf{k}}^{\prime} & h_{12,\mathbf{k}}^{2,S} \\
   H.c. & & & -h_{z,-\mathbf{k}}^{\prime}
  \end{array}\right),
\end{equation}
$h_{2,0}^{S}=3JS\cos2\theta+K_{z}S(2\sin^{2}\theta-\frac{1}{2}\cos^{2}\theta)+h\sin\theta$, $h_{z,\mathbf{k}}^{\prime}=\sin\theta h_{z,\mathbf{k}}$,
$h_{12,\mathbf{k}}^{2,S}=\sin^{2}\theta JSf_{\mathbf{k}}$, $\Delta_{11}^{S}=-\cos^{2}\theta\frac{K_{z}S}{2}$,
$\Delta_{12,\mathbf{k}}^{S}=-\cos^{2}\theta JSf_{\mathbf{k}}$, $\sin\theta=\frac{h}{2S(3J-K_{z})}$. The typical magnon band
in this magnetic phase is shown in Fig.~\ref{fig3} (b), different from the ones in the AFM phase. The first band is always gapless and linearly dispersed
in the vicinity of $\Gamma$ point. The group velocity is very large. For a typical parameters $J=2$ meV, $S=\frac{5}{2}$,
nearest magnetic atom distance $a=0.35$ nm, $K_{z}=0.4$ meV, the group velocity is about $v_{g}\sim5$ km/s.
The low-energy magnons in the SF phase can be used as the information carriers, holding great potential in high-speed information transportation and processing
in the magnonics~\cite{HFu,AVChumak,VVKruglyak,BLenk}.
The gap between the first and second band depends on $h$ and DMI strength $D$. We also labelled the Chern number of the bands.
Similar to the first system, the Hamiltonian in the field-polarized phase is the situation when $\theta=\pi/2$ by leaving the particle branch.

From the two typical magnon spin Hall systems, we can see they exhibit different topological phases in different magnetic ordering.
The Berry curvature distributions also vary for different system and magnetic ordering. Next we will prove that magnon transport properties,
i.e., the Nernst conductivity and thermal Hall conductivity, are similar for the two systems, but different in the three magnetic phases.
As a result, we can use the Nernst effect and thermal Hall effect of magnons to distinguish the magnetic phases experimentally
for magnon spin Hall systems.

\subsection{Magnon Nernst effect and thermal Hall effect}

\begin{figure*}[t]
  \centering
  \includegraphics[width=1.0\textwidth]{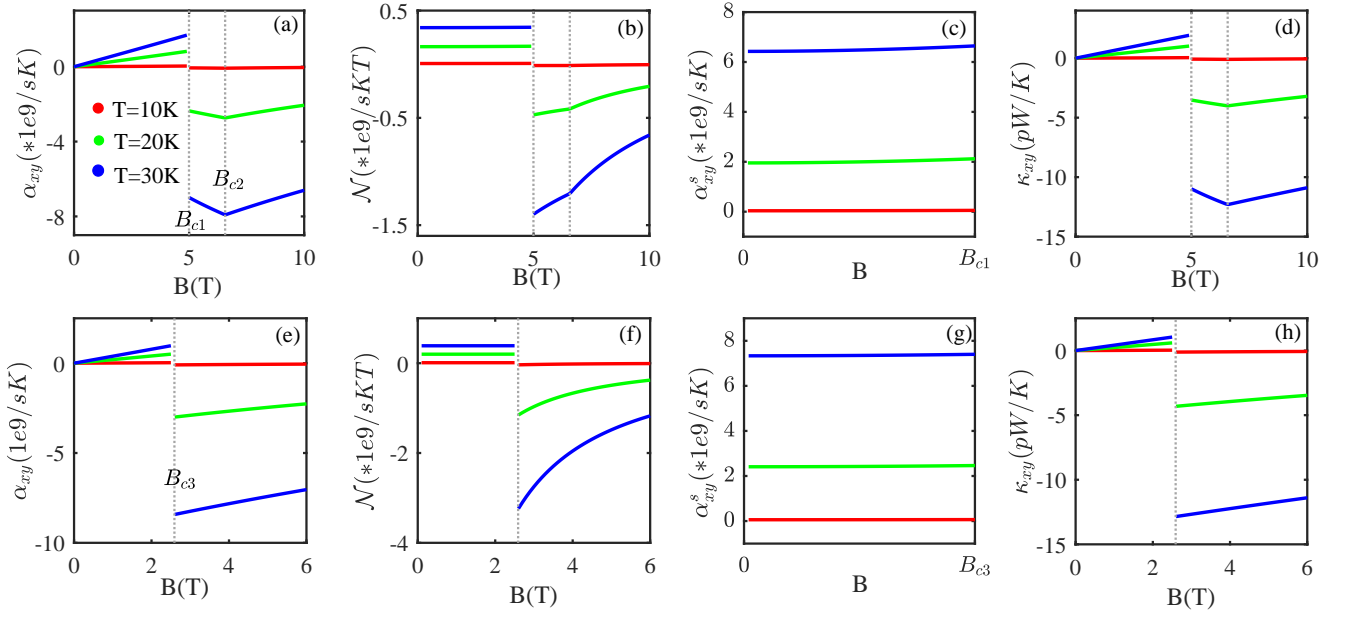}\\
  \caption{The magnon Nernst conductivity ((a) and (e)), Nernst coefficient ((b) and (f)), spin Nernst conductivity ((c) and (g)) and thermal Hall
  conductivity ((d) and (h)) with respect to the magnetic field  for the first magnon spin Hall systems
  at different intelayer coupling strength $J^{\prime}=0.3$ meV ((a)-(d)) and $J^{\prime}=0.15$ meV ((e)-(h)), which give different magnetic phase transition path under magnetic field. The other parameters are adopted the same to Fig.~2 ((b)-(d)).The colors denote different temperatures.}\label{fig4}
\end{figure*}

We now turn to the thermal transport properties induced by the magnons for the magnon spin Hall systems under magnetic field.
The magnetic field lifts the band degeneracy, thus induces imbalanced populations of magnons with opposite Berry curvatures at finite temperatures.
The total effect of Berry curvature does not vanish, and thus manifests Nernst effect and thermal Hall effect under thermal gradient.
We below discuss in details for two systems case by case and summarize the same behaviors.

We first concentrate on the bilayer honeycomb ferromagnets. The weak interlayer AFM coupling strength is comparable
to the easy-axis anisotropy term. There are two possible phase transition paths.
When $K_{z}<J^{\prime}$, the ground state magnetic ordering is from the AFM ordering to the
FP phase via a SF transition. In the AFM ordering, the magnon bands are spin-polarized.
The bands with opposite spin hold opposite Berry curvature at a fixed $\mathbf{k}$ for valence and conduction bands,
$\Omega_{v(c)\uparrow}^{z}=-\Omega_{v(c)\downarrow}^{z}$. At zero magnetic field, it will give us a spin Nernst effect,
vanishing Nernst and thermal Hall effect as $\alpha_{xy,\uparrow}=-\alpha_{xy,\downarrow}$ cancels each other.
At finite field, the spin-polarized bands are separated energetically, inducing an imbalanced spin population,
thus we have $\alpha_{xy,\uparrow}\neq -\alpha_{xy,\downarrow}$. A net Nernst conductivity
$\alpha_{xy}=\alpha_{xy,\uparrow}+\alpha_{xy,\downarrow}$ survives, as shown in Fig.~\ref{fig4} (a).
Here we should notice that the spin Nernst coefficient $\alpha_{xy}^{s}=\alpha_{xy,\uparrow}-\alpha_{xy,\downarrow}$
does not vanish, as shown in Fig.~\ref{fig4} (c). From Fig.~\ref{fig4} (a), we find the Nernst conductivity $\alpha_{xy}$
are linearly dependent on the magnetic field, which leads to a constant Nernst coefficient $\mathcal{N}=\alpha_{xy}/B$
in the AFM phase, as shown in Fig.~\ref{fig4} (b). The Nernst conductivity and coefficient depend on the temperature.

When $B_{c1}<B<B_{c2}$, the ground state magnetic ordering is in the SF phase. Across $B_{c1}$, the system
experience a first-order phase transition and the magnetization change discontinuously.
Correspondingly, the Nernst conductivity $\alpha_{xy}$ shows a discontinuous transition and reverses the sign,
as shown in Fig.~\ref{fig4} (a). The sign reverse is because
the lowest (highest) two magnon bands hold the same sign, different from the AFM phase.
The increasing magnetic field increases the absolute value of $\alpha_{xy}$, which
reaches the extreme at $B=B_{c2}$, shown in Fig.~\ref{fig4} (a).
Further increase of the magnetic field, the spin configuration will go into the FP phase via a second-order phase transition.
From the Nernst conductivity, we can also see
a continuous but not smoothly change, also for the Nernst coefficient $\mathcal{N}$, consistent
with the order of phase transition. Magnons can carry heat, a Nernst effect indicates
a thermal Hall effect. We plot the thermal Hall conductivity $\kappa_{xy}$ with respect to the magnetic field
in Fig.~\ref{fig4} (d). We can see $\kappa_{xy}$ behave the same the $\alpha_{xy}$.
$\kappa_{xy}$ shows a linear behavior in the AFM phase and then a first-order discontinuous
change to the SF phase, then increases to the extreme value and decreases in the FP phase.
For the two magnetic phase transitions,
we can see the transition of the Nernst conductivity and thermal Hall conductivity
is consistent with order parameter transition.
Thus we can use the Nernst conductivity and the thermal Hall conductivity
as the indicator of magnetic phase transitions.

When $K_{z}>J^{\prime}$, the spin configuration will transit form the AFM to the FP phase directly
via a first-order phase transition. In the AFM phase, the Nernst conductivity, Nernst corfficient,
spin Nernst coefficient, and also the thermal Hall conductivity are expected to show same behavior
to the first path, as shown in Fig.~\ref{fig4} (e-h). As there is no SF phase in this path,
the behavior at $B>B_{c3}$ is also the same to the FP phase in the above path.
$\alpha_{xy}$ and $\kappa_{xy}$ show different behavior in the SF and FP phase. If $K_{z}<J^{\prime}$,
We can use the two critical value of magnetic field to find the value of $K_{z}$ and $J^{\prime}$. While
$K_{z}>J^{\prime}$, we can know the $J^{\prime}$ via $B_{c3}$. Here we take the
bilayer CrI$_{3}$ as the example. The neutron scattering experiments can not fully determine
magnetic interaction parameters from the fitting result of the bulk materials experiments~\cite{LChen1,LChen2}.
Our results provide an alternative way to determine the parameters.

\begin{figure}[t]
  \centering
  \includegraphics[width=0.5\textwidth]{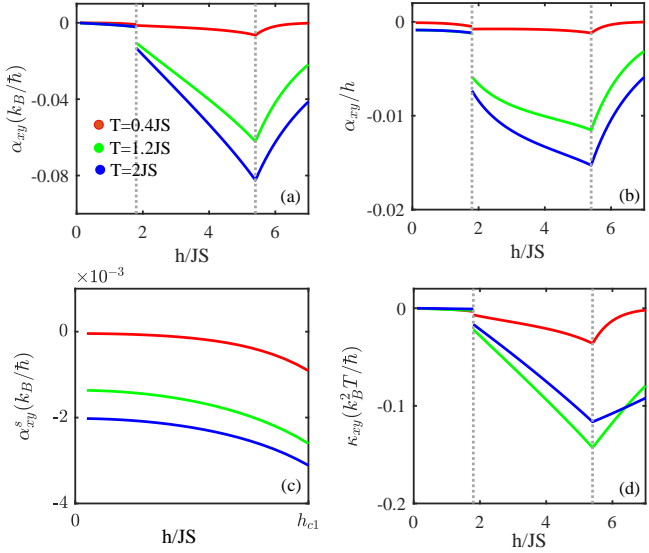}\\
  \caption{The Nernst conductivity (a), the Nernst coefficient (b), the spin Nernst conductivity (c) and thermal Hall conductivity (d) with respect to
  magnetic field for the second spin Hall systems. $D=0.1J$, $K_{z}=0.3J$.}\label{fig5}
\end{figure}

Now we discuss the monolayer honeycomb antiferromagnets.
The easy-axis anisotropy is much smaller than the AFM coupling, $K_{z}<\frac{3}{2}J$.
The ground state spin configuration will go into the FP ordering via a SF transition under magnetic field.
In the AFM ordering, the spin-polarized bands are shifted upwards or downwards by magnetic field.
For each band, we have $\Omega_{\mathbf{k}}^{z}=-\Omega_{-\mathbf{k}}^{z}$,
but $E_{\mathbf{k}}\neq E_{-\mathbf{k}}$ due to the DMI. The nonreciprocal bands hold net Nernst effect
under thermal gradient. But for opposite spins,  $\Omega_{\mathbf{k},\uparrow}^{z}=-\Omega_{\mathbf{k},\downarrow}^{z}$,
the Berry curvature is opposite at fixed $\mathbf{k}$. A zero field give a spin Nernst effect. At finite field,
the split bands give an imbalanced population for opposite spins. The Nernst conductivity is finite but small, shown in
Fig.~\ref{fig5} (a). The same to the first system, the spin Nernst effect survives, with the spin Nernst conductivity
shown in Fig.~\ref{fig5} (c). Here differently, the Nernst conductivity is not linearly dependent on the magnetic field and
is negative-valued.
Across $B_{c1}$, the Nernst and thermal Hall conductivity show a first-order-like phase transition.
In the SF phase, the increase of the magnetic field increases the absolute value of $\alpha_{xy}$ and $\kappa_{xy}$, which
reach the maximum absolute value at $B_{c2}$, same to the first system. Across $B_{c2}$, the $\alpha_{xy}$ and $\kappa_{xy}$
decrease as magnetic field increases, the transition is consistent with the phase transition of order parameter.
Also, we can use the Nernst and thermal Hall conductivity to find the value of $J$ and $K_{z}$.

Above, we have demonstrated theoretically that the magnon spin Hall systems share the same magnetic phase transition diagram by considering
two typical systems.
The Nernst effect, also thermal Hall effect under thermal gradient show distinct behaviors in three magnetic phases but almost the same
for the two systems.
We can use the results of Nernst and thermal Hall conductivity as the indicator of magnetic phase transition and
help us to determine the parameters of anisotropic term and AFM coupling.
The other magnon spin Hall systems are expected to show similar results. In Ref.~\cite{HKondo1}, the bilayer kagome ferromagnets
with AFM interlayer coupling is also a magnon spin Hall system. The behavior of this system under magnetic
field are expected similar to the bilayer honeycomb ferromagnets.
Ref.~\cite{HKondo1,VAZyuzin} present another magnon spin Hall systems, the bilayer honeycomb antiferromagnets
with AFM interlayer coupling. For AA stacking, the interaction energy per unit cell is given by
$\overline{E}=\overline{E}_{intra}+\overline{E}_{inter}+\overline{E}_{Z}+\overline{E}_{ani}$ with
\begin{eqnarray*}
  \overline{E}_{intra}&=&3JS^{2}\cos(\theta_{1A}-\theta_{1B})+3JS^{2}\cos(\theta_{2A}-\theta_{2B}), \\
  \overline{E}_{inter}&=&J^{\prime}S^{2}\cos(\theta_{1A}-\theta_{2A})+J^{\prime}S^{2}\cos(\theta_{1B}-\theta_{2B}), \\
   \overline{E}_{Z} &=&-hS(\sin\theta_{1A}+\sin\theta_{1B}+\sin\theta_{2A}+\sin\theta_{2B}),  \\
  \overline{E}_{ani} &=&-K_{z}S^{2}(\sin^{2}\theta_{1A}+\sin^{2}\theta_{1B}+\sin^{2}\theta_{2A}+\sin^{2}\theta_{2B}).
\end{eqnarray*}
Here the number denotes the layer index and A(B) denote the sublattice. $J$ and $J^{\prime}$ denote the intralayer and interlayer AFM coupling, respectively.
We can see that the interchanges $\theta_{1A}\leftrightarrow\theta_{2B}$, $\theta_{1B}\leftrightarrow\theta_{2A}$ will not change
the interaction energy. Thus the relation $\theta_{1A}=\theta_{2B}$, $\theta_{1B}=\theta_{2A}$ is valid for all magnetic phases. This system is equivalent
to our second system by replacing $J$ with $J+\frac{J^{\prime}}{3}$ for the interaction energy by considering two isolated layers.
Thus the magnetic phase diagram is similar. There are four magnon bands, the topology can be characterized by the spin Chern number in the AFM phase~\cite{VAZyuzin} and Chern number in other two magnetic phases. As a result, the Nernst conductivity and thermal Hall conductivity are expected to behave similarly to the first system under magnetic field.

\section{Conclusions}

In conclusion, we theoretically find the magnon spin Hall system can exhibit Nernst effect and thermal Hall effect under perpendicular magnetic field.
There are three magnetic phases, i.e., the AFM, SF and FP phases. In these phases, the two systems we considered both exhibit Nernst effect
and thermal Hall effect, which show different dependence on magnetic field in the three phases. The transition behavior between the magnetic phases
is consistent with the order of magnetic phase transition, indicating we can use the Nernst effect and thermal Hall effect as
the indicator of magnetic phase transitions, also helping us to determine the magnetic interaction parameters.
Although we only considered two models, the other magnon spin Hall systems can be mapped to the two systems
in the ground state spin configuration and also the thermal transport properties, making our results universal
for magnon spin Hall systems.

\section{Acknowledgements}

Y.-M. Li thanks Dr. Weinan Lin for his helpful discussions. This work is supported by the startup funding from Xiamen University.


\begin{thebibliography}{99}

\bibitem{DXiao1} D. Xiao, M.-C. Chang, and Q. Niu,
Berry phase effects on electronic properties,
\href{https://doi.org/10.1103/RevModPhys.82.1959}{Rev. Mod. Phys. \textbf{82}, 1959 (2010)}.

\bibitem{NNagaosa} N. Nagaosa, J. Sinova, S. Onoda, A. H. MacDonald, and N. P. Ong,
Anomalous Hall effect,
\href{https://doi.org/10.1103/RevModPhys.82.1539}{Rev. Mod. Phys. \textbf{82}, 1539 (2010)}.

\bibitem{JSinova} J. Sinova, S. O. Valenzuela, J. Wunderlich, C. H. Back, and T. Jungwirth,
Spin Hall effects,
\href{https://doi.org/10.1103/RevModPhys.87.1213}{Rev. Mod. Phys. \textbf{87}, 1213 (2015)}.

\bibitem{DXiao3} D. Xiao, W. Yao, and Q. Niu,
Valley-Contrasting Physics in Graphene: Magnetic Moment and Topological Transport,
\href{https://doi.org/10.1103/PhysRevLett.99.236809}{Phys. Rev. Lett. \textbf{99}, 236809 (2007)}.

\bibitem{DXiao2} D. Xiao, G.-B. Liu, W. Feng, X. Xu, and W. Yao,
Coupled Spin and Valley Physics in Monolayers of MoS$_{2}$ and Other Group-VI Dichalcogenides,
\href{https://doi.org/10.1103/PhysRevLett.108.196802}{Phys. Rev. Lett. \textbf{108}, 196802 (2012)}.

\bibitem{JRSchaibley} J. R. Schaibley, H. Yu, G. Clark, P. Rivera, J. S. Ross, K. L. Seyler, W. Yao, and X. Xu,
Valleytronics in 2D materials,
\href{https://doi.org/10.1038/natrevmats.2016.55}{Nat. Rev. Mater. \textbf{1}, 16055 (2016)}.


\bibitem{LZhang} L. Zhang, J. Ren, J.-S. Wang, and B. Li,
Topological Nature of the Phonon Hall Effect,
\href{https://doi.org/10.1103/PhysRevLett.105.225901}{Phys. Rev. Lett. \textbf{105}, 225901 (2010)}.

\bibitem{TQin} T. Qin, J. Zhou, and J. Shi,
Berry curvature and the phonon Hall effect,
\href{https://doi.org/10.1103/PhysRevB.86.104305}{Phys. Rev. B \textbf{86}, 104305 (2012)}.

\bibitem{TSaito} T. Saito, K. Misaki, H. Ishizuka, and N. Nagaosa,
Berry Phase of Phonons and Thermal Hall Effect in Nonmagnetic Insulators,
\href{https://doi.org/10.1103/PhysRevLett.123.255901}{Phys. Rev. Lett. \textbf{123}, 255901 (2019)}.


\bibitem{RYChiao} R. Y. Chiao and Y.-S. Wu,
Manifestations of Berry's Topological Phase for the Photon,
\href{https://doi.org/10.1103/PhysRevLett.57.933}{Phys. Rev. Lett. \textbf{57}, 933 (1986)}.

\bibitem{FDMHaldane} F. D. M. Haldane and S. Raghu,
Possible Realization of Directional Optical Waveguides in Photonic Crystals with Broken Time-Reversal Symmetry,
\href{https://doi.org/10.1103/PhysRevLett.100.013904}{Phys. Rev. Lett. \textbf{100}, 013904 (2008)}.

\bibitem{ZWang} Z. Wang, Y. D. Chong, J. D. Joannopoulos, and M. Solja\v{c}i\'{c},
Reflection-Free One-Way Edge Modes in a Gyromagnetic Photonic Crystal,
\href{https://doi.org/10.1103/PhysRevLett.100.013905}{Phys. Rev. Lett. \textbf{100}, 013905 (2008)}.



\bibitem{RMatsumoto} R. Matsumoto and S. Murakami,
Theoretical Prediction of a Rotating Magnon Wave Packet in Ferromagnets,
\href{https://doi.org/10.1103/PhysRevLett.106.197202}{Phys. Rev. Lett. \textbf{106}, 197202 (2011)}.

\bibitem{HKatsura} H. Katsura, N. Nagaosa, and P. A. Lee,
Theory of the Thermal Hall Effect in Quantum Magnets
\href{https://doi.org/10.1103/PhysRevLett.104.066403}{Phys. Rev. Lett. \textbf{104}, 066403 (2010)}.

\bibitem{YOnose} Y. Onose, T. Ideue, H. Katsura Y. Shiomi, N. Nagaosa,Y. Tokura,
Observation of the Magnon Hall Effect,
\href{https://doi.org/10.1126/science.1188260}{Science \textbf{329}, 297 (2010)}.

\bibitem{AMook1} A. Mook, J. Henk, and I. Mertig,
Magnon Hall effect and topology in kagome lattices: A theoretical investigation,
\href{https://doi.org/10.1103/PhysRevB.89.134409}{Phys. Rev. B \textbf{89}, 134409 (2014)}.

\bibitem{AMook3} A. Mook, J. Henk, and I. Mertig,
Thermal Hall effect in noncollinear coplanar insulating antiferromagnets,
\href{https://doi.org/10.1103/PhysRevB.99.014427}{Phys. Rev. B \textbf{99}, 014427 (2019)}.

\bibitem{YSLu} Y.-S. Lu, J.-L. Li, and C.-T. Wu,
Topological Phase Transitions of Dirac Magnons in Honeycomb Ferromagnets,
\href{https://doi.org/10.1103/PhysRevLett.127.217202}{Phys. Rev. Lett. \textbf{127}, 217202 (2021)}.


\bibitem{RMatsumoto2} R. Matsumoto, R. Shindou, and S. Murakami,
Thermal Hall effect of magnons in magnets with dipolar interaction,
\href{https://doi.org/10.1103/PhysRevB.89.054420}{Phys. Rev. B \textbf{89}, 054420 (2014)}.

\bibitem{PMGunnink} P. M. Gunnink, R. A. Duine, and A. R\"{u}ckriegel,
Theory for electrical detection of the magnon Hall effect induced by dipolar interactions,
\href{https://doi.org/10.1103/PhysRevB.103.214426}{Phys. Rev. B \textbf{103}, 214426 (2021)}.

\bibitem{PMellado} P. Mellado, Intrinsic topological magnons in arrays of magnetic dipoles,
\href{https://doi.org/10.1038/s41598-022-05469-4}{Sci. Rep. \textbf{12}, 1420 (2022)}.

\bibitem{AOkamoto} A. Okamoto and S. Murakami,
Berry curvature for magnons in ferromagnetic films with dipole-exchange interactions,
\href{https://doi.org/10.1103/PhysRevB.96.174437}{Phys. Rev. B \textbf{96}, 174437 (2017)}.



\bibitem{SFujimoto} S. Fujimoto, Hall Effect of Spin Waves in Frustrated Magnets,
\href{https://doi.org/10.1103/PhysRevLett.103.047203}{Phys. Rev. Lett. \textbf{103}, 047203 (2009)}.


\bibitem{KAHoogdalem} K. A. van Hoogdalem, Y. Tserkovnyak, and D. Loss,
Magnetic texture-induced thermal Hall effects,
\href{https://doi.org/10.1103/PhysRevB.87.024402}{Phys. Rev. B \textbf{87}, 024402 (2013)}.

\bibitem{SAOwerre} S. A. Owerre,
Magnon Hall effect without Dzyaloshinskii-Moriya interaction,
\href{https://doi.org/10.1088/0953-8984/29/3/03LT01}{J. Phys.: Condens. Matter \textbf{29} 03LT01 (2017)}.

\bibitem{RRNeumann} R. R. Neumann, A. Mook, J. Henk, and I. Mertig,
Thermal Hall Effect of Magnons in Collinear Antiferromagnetic Insulators: Signatures of Magnetic and Topological Phase Transitions,
\href{https://doi.org/10.1103/PhysRevLett.128.117201}{Phys. Rev. Lett. \textbf{128}, 117201 (2022)}.

\bibitem{RCheng} R. Cheng, S. Okamoto, and D. Xiao,
Spin Nernst Effect of Magnons in Collinear Antiferromagnets,
\href{https://doi.org/10.1103/PhysRevLett.117.217202}{Phys. Rev. Lett. \textbf{117}, 217202 (2016)}.

\bibitem{VAZyuzin} V. A. Zyuzin and A. A. Kovalev,
Magnon Spin Nernst Effect in Antiferromagnets,
\href{https://doi.org/10.1103/PhysRevLett.117.217203}{Phys. Rev. Lett. \textbf{117}, 217203 (2016)}.


\bibitem{YMLi2} Y.-M. Li, J. Xiao, and K. Chang,
Topological Magnon Modes in Patterned Ferrimagnetic Insulator Thin Films,
\href{https://doi.org/10.1021/acs.nanolett.8b00492}{Nano Lett. \textbf{18}, 3032 (2018)}.



\bibitem{HKondo1} H. Kondo, Y. Akagi, and H. Katsura,
$\mathcal{Z}_{2}$ topological invariant for magnon spin Hall systems,
\href{https://doi.org/10.1103/PhysRevB.99.041110}{Phys. Rev. B \textbf{99}, 041110(R) (2019)}.

\bibitem{YMLi} Y.-M. Li, Y.-J. Wu, X.-W. Luo,Y. Huang, and K. Chang,
Higher-order Topological Phases of Magnons in van der Waals Honeycomb Ferromagnets,
\href{https://doi.org/10.48550/arXiv.2202.08424}{arXiv:2202.08424}.



\bibitem{BHuang} B. Huang, G. Clark, E. Navarro-Moratalla, D. R. Klein, R. Cheng, K. L. Seyler, D. Zhong, E. Schmidgall, M. A. McGuire, D. H. Cobden,
W. Yao, D. Xiao, P. Jarillo-Herrero, and X. Xu, Layer-dependent ferromagnetism in a van der Waals crystal down to the monolayer limit,
\href{https://www.nature.com/articles/nature22391}{Nature (London) \textbf{546}, 270 (2017)}.


\bibitem{NSivadas} N. Sivadas, S. Okamoto, X. Xu, Craig. J. Fennie, and D. Xiao,
Stacking-Dependent Magnetism in Bilayer CrI$_{3}$,
\href{https://doi.org/10.1021/acs.nanolett.8b03321}{Nano Lett. \textbf{18}, 7658 (2018)}.


\bibitem{PJiang} P. Jiang, C. Wang, D. Chen, Z. Zhong, Z. Yuan, Z.-Y. Lu, and W. Ji,
Stacking tunable interlayer magnetism in bilayer CrI$_{3}$,
\href{https://doi.org/10.1103/PhysRevB.99.144401}{Phys. Rev. B \textbf{99}, 144401 (2019)}.





\bibitem{AMook2} A. Mook, K. Plekhanov, J. Klinovaja, and D. Loss,
Interaction-Stabilized Topological Magnon Insulator in Ferromagnets,
\href{https://doi.org/10.1103/PhysRevX.11.021061}{Phys. Rev. X \textbf{11}, 021061 (2021)}.

\bibitem{RShindou} R. Shindou, R. Matsumoto, S. Murakami, and J.-ichiro Ohe,
Topological chiral magnonic edge mode in a magnonic crystal,
\href{https://doi.org/10.1103/PhysRevB.87.174427}{Phys. Rev. B \textbf{87}, 174427 (2013)}.


\bibitem{LChen1} L. Chen, J.-H. Chung, B. Gao, T. Chen, M. B. Stone, A. I. Kolesnikov, Q. Huang, and P. Dai,
Topological Spin Excitations in Honeycomb Ferromagnet CrI$_{3}$,
\href{https://doi.org/10.1103/PhysRevX.8.041028}{Phys. Rev. X \textbf{8}, 041028 (2018)}.

\bibitem{LChen2} L. Chen, J.-H. Chung, M. B. Stone, A. I. Kolesnikov, B. Winn, V. O. Garlea, D. L. Abernathy, B. Gao, M.Augustin, E. J. G. Santos, and P. Dai,
Magnetic Field Effect on Topological Spin Excitations in CrI$_{3}$,
\href{https://doi.org/10.1103/PhysRevX.11.031047}{Phys. Rev. X \textbf{11}, 031047 (2021)}.



\bibitem{LSheng} L. Sheng, D. N. Sheng, C. S. Ting, and F. D. M. Haldane,
Nondissipative Spin Hall Effect via Quantized Edge Transport,
\href{https://doi.org/10.1103/PhysRevLett.95.136602}{Phys. Rev. Lett. \textbf{95}, 136602 (2005)}.

\bibitem{DNSheng} D. N. Sheng, Z. Y. Weng, L. Sheng, and F. D. M. Haldane,
Quantum Spin-Hall Effect and Topologically Invariant Chern Numbers,
\href{https://doi.org/10.1103/PhysRevLett.97.036808}{Phys. Rev. Lett. \textbf{97}, 036808 (2006)}.

\bibitem{EProdan} E. Prodan, Robustness of the spin-Chern number,
\href{https://doi.org/10.1103/PhysRevB.80.125327}{Phys. Rev. B \textbf{80}, 125327 (2009)}.

\bibitem{YYang} Y. Yang, Z. Xu, L. Sheng, B. Wang, D. Y. Xing, and D. N. Sheng,
Time-Reversal-Symmetry-Broken Quantum Spin Hall Effect,
\href{https://doi.org/10.1103/PhysRevLett.107.066602}{Phys. Rev. Lett. \textbf{107}, 066602 (2011)}.




\bibitem{AVChumak} A. V. Chumak, V. I. Vasyuchka, A. A. Sergam, and B. Hillebrands,
Magnon spintronics,
\href{https://doi.org/10.1038/nphys3347}{Nature Phys. \textbf{11}, 453 (2015)}.

\bibitem{VVKruglyak} V. V. Kruglyak, S. O. Demokritov, and D. Grundler, Magnonics,
\href{http://dx.doi.org/10.1088/0022-3727/43/26/264001}{J. Phys. D: Appl. Phys. \textbf{43}, 264001 (2010)}.

\bibitem{BLenk} B. Lenk, H. Ulrichs, F. Garbs, M. M\"{u}nzenberg, The building blocks of magnonics,
\href{https://doi.org/10.1016/j.physrep.2011.06.003}{Phys. Rep. \textbf{507}, 107 (2011)}.


\bibitem{HFu} H. Fu, K. Huang, K. Watanabe, T. Taniguchi, and J. Zhu,
Gapless Spin Wave Transport through a Quantum Canted Antiferromagnet,
\href{https://doi.org/10.1103/PhysRevX.11.021012}{Phys. Rev. X \textbf{11}, 021012 (2021)}.




\end{thebibliography}
\end{document}